\documentclass[%
 reprint,
 amsmath,amssymb,
 aps,
]{revtex4-1}

\usepackage{graphicx}
\usepackage{dcolumn}
\usepackage{bm}

\begin{document}


\title{Numerical study of variable lung ventilation strategies}
\thanks{Published in Lecture Notes in Electrical Engineering, Springer, 2015}%

\author{Reena Yadav}
\affiliation{Centre for Information Communication Technology, Indian Institute of Technology Jodhpur, India}%

\author{Mayur Ghatge}%
\affiliation{Centre for Information Communication Technology, Indian Institute of Technology Jodhpur, India}%

\author{Kirankumar Hiremath}
\affiliation{Centre for System Science, Indian Institute of Technology Jodhpur, India}%

\author{Ganesh Bagler}
\homepage{home.iitj.ac.in/$\sim$bagler/}
\email{bagler@iitj.ac.in}
\affiliation{Centre for Biologically Inspired System Science, Indian Institute of Technology Jodhpur, India}%


\begin{abstract}
Mechanical ventilation is used for patients with a variety
of lung diseases. Traditionally, ventilators have been designed to
monotonously deliver equal sized breaths. While it may seem intuitive
that lungs may benefit from unvarying and stable ventilation pressure
strategy, recently it has been reported that variable lung ventilation
is advantageous. In this study, we analyze the mean tidal volume in
response to different `variable ventilation pressure' strategies. We 
found that uniformly distributed variability in pressure gives the best tidal volume as compared to that of normal, scale-free, log normal and linear distributions.
 
\begin{description}
\item[Cite as:]
R Yadav, M Ghatge, K Hiremath and G Bagler, Chapter 26, pp 299-306, Systems Thinking Approach for Social Problems, Lecture Notes in Electrical Engineering, 327, 2015.
\end{description}
\end{abstract}

\pacs{Valid PACS appear here}
\maketitle



\section{\label{sec:intro}INTRODUCTION}
Biological systems are inherently adaptive and have evolved to survive stressors and randomness. Beyond being robust to stressors, they are in fact evolutionarily designed to `gain' from them. This property of systems has been referred to as `antifragility' \cite{2012arXiv1208.1189T,taleb2012antifragile}. Antifragile systems benefit from stressors or noise. Here, by `stressors' one refers to unfavorable abiotic changes in external milieu such as temperature,  pH, pressure etc.  A system's reaction to any external stressor or stimuli could be enumerated through a measurable property that reflects gain or loss in response to the stressor. It has been argued that a system having convex response would benefit from addition of noise~\cite{2012arXiv1208.1189T,taleb2012antifragile}. Many of the biological systems have convex responses due to evolutionary selection that they have undergone. Modeling the nature of such system responses would help in better understanding of design principles of biological systems that help them to thrive under stressful circumstances.

Human engineered systems are designed to cope with stable signals. An electrical system is designed assuming unvarying electrical signal and structural systems are designed for absence of severe seismic disturbances. Thus, signal variability reflects `noise' and is harmful to the system. Classically physiological systems are thought to be designed to reduce variability and to attain homeostasis. In contrast, signals of a wide variety of physiological systems, such as human heartbeat, brain's electrical activity etc., fluctuate in a complex manner. 
In fact, a defining feature of a living organic system is adaptability, the capacity to respond to unpredictable stimuli. 

\begin{figure*}[htb!]
	\includegraphics[width=\textwidth]{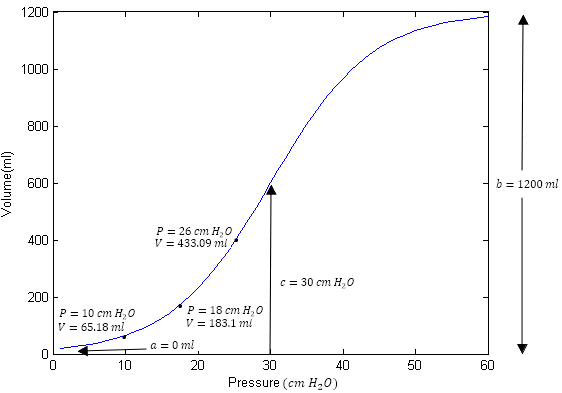}
	\caption{\label{scc}The generic static compliance curve (Pressure-Volume response curve) obtained from experimental data collected on mechanically ventilated lungs. The curve is nonlinear and roughly sigmoidal in shape~\cite{venegas1998comprehensive}. Various parameters used in Eqn.\ \eqref{eqscc} have values as shown in the figure.}
\end{figure*}

A lung is a physiological system that serves towards respiration, critical for oxidative processes in organisms. Human breathing is driven by pressure generated through spontaneous muscular action. The lungs are ventilated in response to this pressure. The lung volume representing the normal volume of air displaced between normal inspiration and expiration, when no extra effort is applied, is referred to as `tidal volume'. The pressure-volume response curves for lung are known as `static compliance curve'. Figure~\ref{scc} shows the standard static compliance curve for normal human lung ventilation~\cite{venegas1998comprehensive}. 
Mathematical modeling of this process results in a sigmoidal equation of the form:
\begin{equation}
	v= a+ b \frac{1}{1+ \exp^{-((p-c)/d)}}\quad\text{,} 
	\label{eqscc}
\end{equation}
where $v$ is volume and $p$ is pressure.\\

\indent The equation has been shown to best fit the curve for both the convex and concave region~\cite{venegas1998comprehensive}. This equation not only comprehensively characterizes the P-V curve but also provides various parameters essential for clinical experimentation and studies~\cite{venegas1998comprehensive}.
The four parameters given by $a$, $b$, $c$, $d$ in the equation are fitting parameters. The parameter $a$ is the lower asymptote volume; here $0\:ml$ and $b$ corresponds to difference between lower and higher asymptote; here it is $1200\:ml$. The parameter $c$ depicts the pressure at the inflection point which is given by $30\:cm$ of $H_{2}O$. Finally, $d$ is proportional to the pressure range within which most of the volume change takes place i.e.\ it is the index of linear compliance with a value of $7\:cm$ of $H_{2}O$~ \cite{venegas1998comprehensive,brewster2005convexity}.

When lungs are incapable of ventilating spontaneously, mechanical ventilators are used. Historically, mechanical ventilators are designed to deliver equal sized breaths. It has been observed that such conventional monotonously regular ventilation  
has negative consequences for critically ill patients~\cite{ats1999srlf, albert2012role,dos2000invited}. For instance, patients suffering from Acute Respiratory Distress Syndrome (ARDS) can have negative impact from this conventional mode of mechanical ventilation due to alveolar collapse and airway damage~\cite{rouby2007tidal,ats1999srlf,albert2012role}. Natural healthy ventilation is characterized by its variability. Biologically variable ventilation emulates healthy variation and has been shown to prevent deterioration of gas exchange~\cite{mutch2000biologically2}, increase arterial oxygenation~\cite{mutch2000biologically} and, in general, is reported to improve respiratory mechanics under various lung pathologies~\cite{mcmullen2006biologically, boker2004variable,mutch2000biologically1,mutch2007biologically}. The reasons for advantageous effects of variable ventilation are not entirely clear and need to be explored further.

\begin{figure*}[tb!]
	\includegraphics[width=\textwidth]{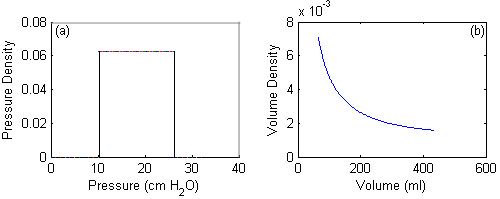}
	\caption{{The nature of tidal volume distribution (b) in response to uniformly distributed variable ventilation pressure (a), as given by Eqn.\ \eqref{ud}.} }
	\label{ud}
\end{figure*}

In this study, we explore whether by adding suitably designed noise in ventilation pressure of mechanical ventilators, one can obtain better tidal volume without increasing the mean airway pressure. 
Apart from ventilation pressure distributed as uniform distribution~\cite{brewster2005convexity}, we studied Gaussian, log-normal, linear and power law distributions.
In contrast to uniform distribution which gives equal weightage to convex and concave parts of the static compliance curve, these distributions preferentially span the curve. This allowed us to focus on different parts of the curve.
For example, Gaussian distribution gives more emphasis on the central part of the curve. The log-normal and linear stress on the latter half of the applied pressure range, whereas power law emphasizes on the first half of pressure range while introducing a few high-pressure spikes. Studying various variable ventilation strategies could provide us an insight into the best possible method for operating mechanical ventilators so as to benefit the most from convexity of the static compliance curve.

Jensen's inequality provides the central argument to explain benefits of convexity and could be used to identify conditions under which addition of noise will be beneficial~\cite{brewster2005convexity}. Jensen's inequality states that if $f(x)$ is a real valued convex function in the interval $[a, b]$ and $X$ is a random variable within the range $[a, b]$ then,
$$  f (E[X]) \le E [f(X)] \quad\text{,}$$
here $E[X]$ is the expected value (i.e.\ mean) of the (random) variable.\\

\indent As a consequence of Jensen's inequality, varying the pressure within the convex region, before the inflection point, of the static compliance curve i.e.\ from 10 cm $ H_{2}O$ to 26 cm $H_{2}O$ (Figure \ref{scc}) would result in higher mean tidal volume in contrast to constant pressure strategy.
The gain in mean tidal volume, as compared to application of constant pressure, depends on the range of static compliance curve spanned and the frequency.



\section{Studies}
We studied various variable ventilation strategies to study their effect on mean tidal volume in contrast with the conventional mode of mechanical ventilation.
`Constant Mode Ventilation' monotonously delivered air into the lungs at regular intervals and at the same pressure. 
In recent studies \cite{brewster2005convexity} it has been shown that uniformly varying pressure values in the range from $10\:cm$ $ H_{2}O$ to $26\:cm$ $H_{2}O$ results in better tidal volume compared to `constant mode ventilation'. Figure \ref{ud}(a) shows the tidal volume output for the uniformly distributed pressure values and the corresponding volume distribution obtained as shown in Figure \ref{ud}(b). The mean tidal volume improves to $205.72\:ml$ from $183.13\: ml$ obtained using the monotonous strategy, even though both strategies use the same mean pressure of $18\:cm$ $H_{2}O$.

The Uniform Pressure Distribution function can be mathematically written as -
\begin{eqnarray}
	\mathcal{U} &=&
	\left\{\begin{array}{ll}
		\frac{1}{16} & \mbox{if }  $10 \! cm$~H_{2}O\le p \le$26 \! cm$~H_{2} O\\
		0 &\mbox{otherwise}
	\end{array}\right.
\end{eqnarray}

Variably distributed pressure values are sampled from the interval  of ($10\:cm$ $ H_{2}O$, $26\:cm$ $H_{2}O$), which is the physiological limit of minimum and maximum pressure values for human lungs \cite{rouby2007tidal}. 
Also as can be seen from the Static Compliance Curve in Figure \ref{scc}, this interval lies in the convex region of the curve covering the output volume
range of $65.18\: ml$ to $433.09\: ml$ as opposed to a single value of $183.13\: ml$ corresponding to $18\: cm$ pressure, used in constant mode ventilation strategy. 

\begin{figure*}[htb!]
	\includegraphics[width=\textwidth]{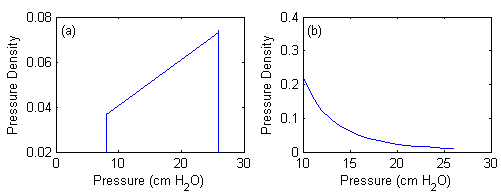}
	\caption{{(a) Linear pressure variation with $8\:cm\le p \le 26\:cm $.  (b) Pressure variation following power law distribution, given by Eqn.\ \eqref{pl} with $\alpha$ = 2.25 and $x_{min}$ = 10 \:cm $H_{2}O$ } }
	\label{lppd}
\end{figure*}

\begin{figure*}[htb!]
	\includegraphics[width=\textwidth]{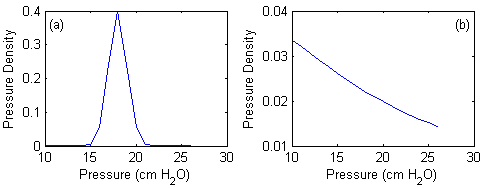}
	\caption{{(a) Gaussian Distribution given by Eqn.\ \eqref{nd} and (b) Log-normal Distribution represented by Eqn.\ \eqref{lgn} with $\mu$= $18\: cm$ $H_{2}O$ and $\sigma$ = 1. } }
	\label{nlgpd}
\end{figure*}

We further explored the effect of variable ventilation in the form of a few canonical distributions on the tidal volume. The mean value of pressure is always kept constant at $18\:cm$ in all the distributions discussed throughout our studies. 
The density function equations for various pressure distributions are as follows. 
\begin{itemize}
	\item
	\textbf{Gaussian Pressure Distribution}
	\begin{equation}\label{nd}
		\mathcal{N} = \frac{1}{{\sigma \sqrt {2\pi } }}e^{{{ - \left( {x - \mu } \right)^2 } \mathord{\left/ {\vphantom {{ - \left( {x - \mu } \right)^2 } {2\sigma ^2 }}} \right. \kern-\nulldelimiterspace} {2\sigma ^2 }}}
	\end{equation}
	where $\mu$ is the mean of the distribution and $\sigma$ represents standard deviation.
	\item
	\textbf{Power-Law Pressure Distribution}
	\begin{equation}\label{pl}
		\mathcal{P} = \frac{\alpha}{x_{min }}(\frac{x}{x_{min} })^{-\alpha-1}
	\end{equation}
	where $x_{min}$ is the minimum possible value of $x$ and $\alpha$ is the power-law exponent with value greater than 1. 
	\item
	\textbf{Log-Normal Pressure Distribution}
	\begin{equation}\label{lgn}
		Log-\mathcal{N} = \frac{1}{{x\sigma \sqrt {2\pi } }}e^{{{ - \left( {ln(x) - \mu } \right)^2 } \mathord{\left/ {\vphantom {{ - \left( {x - \mu } \right)^2 } {2\sigma ^2 }}} \right. \kern-\nulldelimiterspace} {2\sigma ^2 }}}
	\end{equation}
\end{itemize}
where $\mu$ is the mean and $\sigma$ is standard deviation.

Figure~\ref{lppd} depicts the nature of pressure density functions for linear and power law distributions and Figure \ref{nlgpd} depicts that for Gaussian and log normal distributions.

\begin{figure*}[htb!]
	\includegraphics[width=\textwidth]{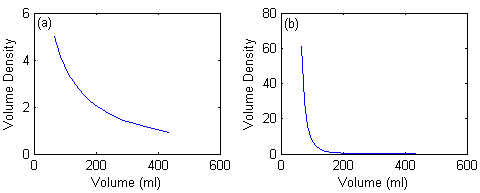}
	\caption{{Tidal Volume output for (a) Linear and (b) Power law pressure distributions.} }
	\label{lpvd}
\end{figure*}

The linear pressure distribution in Figure~\ref{lppd}(a) gives volume density function which is only marginally different from that of uniform distribution. The maximum mean tidal volume produced is $205.59\:ml$ with minimum pressure reduced to $8\:cm$ from that of $10\:cm$ used in all other cases to maintain the average pressure at $18 \: cm $. Thus linear distribution comes close to uniform in terms of the mean output volume. The slope and volume intercept in Figure \ref{lppd}(a) are $1/486$ and $10/486$ respectively. The minimum pressure applied was changed to $5\:cm$ while keeping the maximum pressure at $26\,cm$ $H_{2}O$ to see the effect of change in slope to the volume output. However it was observed that even after changing the minimum pressure value, the mean tidal volume does not vary much from that observed with the pressure of $8\:cm$ (Figure~\ref{lpvd}).

\begin{figure*}[htb!]
	\includegraphics[width=\textwidth]{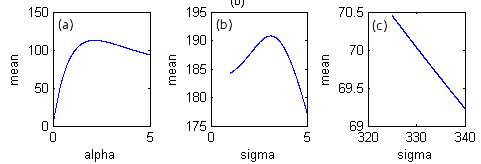}
	\caption{{Effect of parameters $\alpha$ (exponent in Eqn.\  \eqref{nd}) and $\sigma$ (standard deviation in Eqns.\ \eqref{nd} and \eqref{lgn}) on the mean tidal volume output. (a) Power law distribution. (b) Gaussian distribution. (c) Log-normal distribution.} }
	\label{as}
\end{figure*}

The pressure values in Figure~\ref{lppd}(b) follow power law distribution within the same interval ($10\:cm~H_{2}O$,  $26\:cm$ $H_{2}O)$.  The power law density function gives maximum volume output of $112.3\:ml$ for $\sigma$ = 1.93, beyond which the volume starts decreasing as is evident from Figure~\ref{as}(a). This volume output is  worse than that of constant mode ventilation, contrary to the proposal that noisy ventilation should be beneficial in general.
\begin{figure*}[htb!]
	\includegraphics[width=\textwidth]{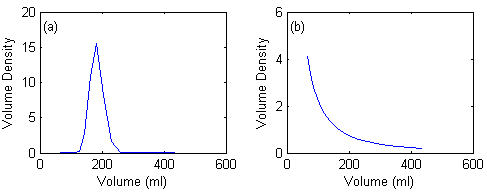}
	\caption{{Tidal volume output for  (a) Gaussian and (b) Log-normal pressure distributions.} }
	\label{nlgvd}
\end{figure*}

When pressure values are sampled from Gaussian distribution as shown in the Figure~\ref{nlgpd}(a), the performance of the volume density function improves from that of constant mode, though not significantly. The mean volume varies with $\sigma$ and has the maximum value of $190.8\:ml$ at $\sigma$ = $3.15$ after which the volume starts decreasing sharply as shown in Figure~\ref{as}(b).

When the probability density function of pressure follows log-normal distribution (Figure~\ref{nlgpd}(b)) the mean tidal volume output was found  to be $70.44\:ml$, worse than even that of `constant mode ventilation'. Thus log-normal strategy resulted in the worst performance among all the five canonical distributions studied. Figure~\ref{as}(c) shows that the maximum value for volume output in the log-normal case comes at 
$\sigma$= $325.1$ after which the mean volume decrease sharply.

\section{CONCLUSIONS}
Variable ventilation strategy in general is reported to be beneficial in terms of mean tidal volume, without increasing the ventilation pressure. It is desirable to characterize the effect of various variable ventilation strategies, which can help in the identification of best possible strategies to implement in mechanical ventilators. From our studies with five canonical distribution strategies, we found that uniform pressure distribution is the most favorable `variable pressure strategy'. This distribution exploits the convexity of the generic static compliance curve the most, to emerge as the best variable ventilation strategy among the canonical distributions studied in this paper.  

This leads to an important question as to whether uniform distribution is the ``best possible'' noisy strategy for the pressure density function or could there be better strategies. Within the constraints of physiologically acceptable range of pressures and without increasing the mean pressure, this question could be posed as an optimization problem. 

For any distribution of pressure values there could be multiple time sequences with which these pressure instances could be applied. It needs to be studied further which of these time sequences could be physiologically meaningful. Also, one could study empirical distributions for various breathing patterns such as normal breathing, panting and \emph{pranayama}.

\end{document}